\documentclass[12pt]{article}
\def\thb{\bar \theta}
\def\eb{\bar\e}

\newfont{\goth}{eufm10 scaled \magstep1}

\def\a{\alpha}

\def\b{\beta}

\def\c{\gamma}\def\C{\Gamma}
\def\d{\delta}
\def\e{\epsilon}

\def\k{\kappa}

\def\m{\mu}
\def\n{\nu}

\def\s{\sigma}
\def\t{\tau}
\def\th{\theta}

\def\beq{\begin{equation}}\def\eeq{\end{equation}}
\def\beqa{\begin{eqnarray}}\def\eeqa{\end{eqnarray}}
\def\barr{\begin{array}}\def\earr{\end{array}}

\def\del{\partial}\def\delb{\bar\partial}

\def\una{{\underline a}}\def\unA{{\underline A}}
\def\unb{{\underline b}}\def\unB{{\underline B}}

\def\unM{{\underline M}}

\def \ys {{y\kern-.5em / \kern.3em}}

\def\Ch{\hat{\C}}

\let\bm=\bibitem

\def\uM{{\underline M}}
\def\umu{{\underline \mu}}
\def\unu{{\underline \nu}}
\def\nn{\nonumber}
\def\bd{\begin{document}}
\def\ed{\end{document}}
\def\ba{\begin{array}}
\def\ea{\end{array}}
\def\bea{\begin{eqnarray}}
\def\eea{\end{eqnarray}}
\def\ft#1#2{{\textstyle{{\scriptstyle #1}\over {\scriptstyle #2}}}}
\def\fft#1#2{{#1 \over #2}}
\newcommand{\be}{\begin{equation}}
\newcommand{\ee}{\end{equation}}
\newcommand{\eq}[1]{(\ref{#1})}
\def\eqs#1#2{(\ref{#1}-\ref{#2})}
\def\det{{\rm det\,}}
\def\tr{{\rm tr}}
\newcommand{\ho}[1]{$\, ^{#1}$}
\newcommand{\hoch}[1]{$\, ^{#1}$}
\def\ra{\rightarrow}
\def\uha{{\hat {\underline{\a}} }}
\def\uhc{{\hat {\underline{\c}} }}

\def \Om {\Omega}
\def \bfd {{\bf d}}
\def \del {\partial}
\def \eps {\epsilon}
\def \Z {{\bf Z}}
\def \xb {\bar{x}}
\def \la {\langle}
\def \ra {\rangle}
\def \Omt {\tilde \Omega} 
\def \la {\langle}
\def \ra {\rangle}

\begin{document}

\hfill{NEIP-019}

\hfill{UTTG-09-99}

\hfill{hep-th/9912153}

\vspace{20pt}

\begin{center}

{\LARGE \bf
Manifest Supersymmetry \\
\vspace{10pt}
in Non-Commutative Geometry}

\vspace{30pt}

{\large Chong-Sun Chu\hoch{1} and  Frederic Zamora \hoch{2}}

\vspace{15pt}

\begin{itemize}
\item[$^1$] {\em Institute of Physics, University of
Neuch\^atel, CH-2000 Neuch\^atel, Switzerland}
\item[$^2$] {\em Theory Group, Department of Physics, 
University of Texas, Austin TX 78712 USA}
\end{itemize}

\vskip .2in \sffamily{chong-sun.chu@iph.unine.ch\\ }
\vskip .1in \sffamily{zamora@zerbina.ph.utexas.edu \\}
\vspace{60pt}

{\bf Abstract}
\end{center}
We consider the open superstring ending on a D-brane in the presence of a 
constant NS-NS $B$ field, using the Green-Schwarz formalism. 
Quantizing in the light-cone gauge, we find that 
the anti-commutation relations for the fermionic variables
of superspace remain unmodified. We also derive the 
unbroken supersymmetry algebra living on the D-brane.
This establishes how the Moyal product is extended in 
a superspace formulation of non-commutative field theories.
The superfield formulation of non-commutative supersymmetric field
theories is briefly considered. 

\newpage


\section{Introduction}

During the last years,
there has been remarkable progress in understanding
the dynamics of supersymmetric gauge theories and string 
theory. In fact, both fields have benefited from each other, in 
elaborate and beautiful ways. 
Recently, a new connection between them has arisen:
string theory backgrounds with a constant 
NS-NS $B$ field correspond, in the Sen-Seiberg limit \cite{SS}, 
to non-commutative gauge theories
\cite{CDS,SW}. The dynamics of these theories appears up to now 
quite mysterious, in particular at distances comparable to 
the non-commutativity scale. Looking back to 
the examples of ordinary supersymmetric gauge theories, 
it seems quite possible that supersymmetry will provide 
strong constraints on the dynamics of these non-commutative 
field theories.

Therefore, it seems worthwhile to analyze how supersymmetry 
is realized in these theories. 
To this end, we would like to work in a 
context where supersymmetry is as manifest as possible.
For certain numbers of spacetime dimensions and supersymmetries,
one can construct a superspace and superfield formalism 
for ordinary supersymmetric field theories. 
It is very likely that a similar construction is available 
for non-commutative supersymmetric field theories,
where some additional fermionic variables promote a given 
manifold to a supermanifold. 
But then, 
as a mathematical question in the context of non-commutative
geometry, one could imagine different consistent 
deformations of the fermionic anti-commutation rules. For example,
a possible deformation of the multiplication 
between fermionic coordinates of superspace could be 
\be
f(\theta)\star g(\theta) = e^{ \frac{\Delta^{\a\b}}{2}
\frac{\del}{\del \th^\a}\frac{\del}{\del \th^{'\b}} } 
f(\theta) g(\theta')|_{\th=\th'} ,
\ee
where $\Delta^{\a\b}$ is symmetric. 
An explicit deformation of $N=1$ superspace  in two
dimensions has been considered in \cite{fz}.
Quantum group \cite{frt} like deformations may also be envisaged.
To find out what happens in string theory is one of the
purposes of this letter.
 
In previous papers \cite{CH1,CH2,Schom,aasj}, the quantization of 
an  open string ending on a D-brane 
with a constant NS-NS $B$-field background 
was carried out in the RNS formulation.
The result was that the spacetime coordinates 
on the world-volume of the D-brane were non-commutative.
In this paper we analyze the same situation, but with 
the string formulated in the Green-Schwarz
\cite{GS}
language\footnote{The analysis of D-branes from 
the Green-Schwarz superstring, 
without a background $B$-field, has been considered in 
\cite{GG,LW}}.
We will take advantage of the manifest spacetime supersymmetry 
in this formalism to obtain the anti-commutation relations 
for the Grassmanian variables $\theta$  of superspace,
in a situation where the bosonic coordinates $X$ do not commute.
The result is that the fermionic coordinates are \emph{undeformed},
and have the same anti-commutation rules on the brane.
Also, we will derive the supersymmetry algebra living on the 
D-brane and verify that its structure is not modified. 
Hence, if we want to work with superfields evaluated on 
flat superspace, their non-commutative associative $*$-product 
reduces to the ordinary anti-commuting product 
for the fermionic variables $\theta$ of superspace. 

The paper is organized as follows. 
We start in section 2, reviewing
the construction of the Green-Schwarz (GS) superstring 
action in a general type II supergravity background; after that, 
we deal with  the case of interest: a flat background with a constant 
NS-NS $B$ field, where the D-brane boundary conditions are established 
and solved. In section 3 we carry out the light-cone quantization 
of the GS superstring; we compute the unbroken supersymmetry algebra
living on the world-volume of the D-brane, finding the same 
structure as in the ordinary (commutative) theories;
we also show how the appearance of the Moyal product can be understood 
from  the oscillator quantization method.
Finally, in section 4 we introduce the notions of
superspace and superfields in non-commutative geometry. 
The WZ model is discussed as an illustrative  example,
and we observe that the perturbative 
non-renormalization theorem for the F-term is still valid.

\bigskip

\section{D-branes in the Covariant Green-Schwarz \\
Formalism}

\subsection{Green-Schwarz Superstring in a general Type II background}

The first thing we need to know is how the 
GS superstring couples to the $B$-field. 
In this subsection we review the construction of \cite{Grisaru},
which ensures a $\kappa$-invariant action.

Let us start by setting up some conventions.
The ten dimensional ${\cal N}=2$ superspace  
is a supermanifold of dimension $(10|32)$. 
Since we will first present the GS superstring propagating 
in a general type II supergravity background, we
denote by $Z^\uM= (X^\umu, \theta^m)$ its local coordinates 
and by $\unA=(\una,\a)$ its target space tangent indices.
Underlining an index indicates that it is 10 dimensional
\footnote{We are following basically the notation and conventions 
of \cite{CHS} for the super-embedding, except
that, in order to simplify the notation, 
we will not distinguish whether a fermionic index is 10
dimensional or $(p+1)$-dimensional (for a D$p$-brane)
and so we will not underline the fermionic indices. This is
sufficient for our purposes, as we will not need to split the
10-dimensional $\C$-matrices into those appropriate for the
$(p+1)$-dimensional submanifold. All
the $\C$-matrices in this paper are the 
ten dimensional ones}. 
The super-vielbeins are denoted by $E_\uM{}^\unA$. 
The fermionic variables $\theta$ are spacetime spinors and 
consist of two Majorana-Weyl spinors. In type IIA
superspace, these spinors carry opposite chiralities, which can be
combined into a single 32 component Majorana spinor; while in type IIB
they are of the same chirality. We will use the fermionic
index $\a$ for both cases. In type IIA, it is a 32 dimensional spinor 
index, while in type IIB it is
understood to be a composite index of a Majorana spinor index (32
dimensional) and an $SO(2)$ doublet index acted on by the 
Pauli matrices. Chiral projection operators 
should be inserted in appropriate places to
reduce the Majorana indices to Majorana-Weyl ones.

The covariant $\kappa$-symmetric
action for the $N=2$ superstring in a supergravity background 
takes the form \cite{Grisaru}
\be \label{action}
I = -\frac{1}{2\pi\a'}\int d^2 \xi\,
\{\sqrt{-g} g^{ij}  E_i{}^\una E_j{}^\unb \eta_{\una\unb} 
- \e^{ij} E_i{}^\unB E_j{}^\unA B_{\unA\unB}\} \, ,
\ee
where $g_{ij}$ is the metric of the world-sheet,
$\e^{01} =1$, 
$ \eta_{ab}={\rm diag}(-,+,...,+)$ is the 10d Minkowski
spacetime metric and the pull-back super-vielbeins are
\be \label{pullE}
E_i{}^\unA \equiv \del_i Z^\unM E_\unM{}^\unA =(E_i{}^\una, E_i{}^{\a}
). 
\ee
What is relevant for us is the second term in \eq{action},
the pull-back of the 2-form superfield
\be
B =\frac{1}{2}  E^\unB E^\unA B_{\unA\unB} \,,
\ee
which is the (local) potential of the closed super 3-form $H=dB$
of type II supergravity.
Since we will consider world-sheets with boundaries, properly
there are also the vector superfields $A$ with supersymmetric 
couplings to the string \cite{CHS}. 
But in order to simplify the discussion,
and without any loss of generality for the case of 
constant $B$-field, we confine to  
the gauge where $A$ is zero.

The action \eq{action} is invariant under
the $\k$-symmetry:
\bea
\d_{\k} Z^{\una} &=& 0 \ ,\nn\\
\d_{\k} Z^{\a} &=&  \ft12\,\k^{\c}(\xi) (1+\C)_{\c}{}^{\a} \ ,
\eea
with the definitions
$ \d_{\k} Z^{\unA} := \d_{\k} Z^{\unM} E_\unM{}^\unA$ and 
\bea 
(\C)_{\a}{}^{\b} &=& \frac1{2\sqrt{-g}}\e^{ij}
\left(E_i{}^\una E_j{}^\unb \C_{\una\unb} P\right)_{\a}{}^{\b}\ ,
\nn\\
 P&=&\cases{\C_{11} & (IIA) \cr
       \s_1& (IIB)} \,.
\eea
Note that $\C_{11}$ acts on the 32-component Majorana index, while
$\s_3$ acts on the $SO(2)$ doublet index of the two 16-component
Majorana-Weyl spinors of type IIB.

The constraints which follow by demanding
$\k$-symmetry invariance of \eq{action} correspond to the ones
of the 10d ${\cal N}=2$ supergravity multiplet \cite{type2}.
There are the constraints on the torsion super two-form,
\bea
\label{T}
T_{\a\b}{}^{\una} &=& -2i(\C^{\una})_{\a\b}\ , \nn\\
T_{\a \unb}{}^{\una} &=& \d_{\unb}{}^{\una}~\chi_{\a}\ , 
\eea
and on the super three-form,
\bea \label{H}
&& H_{\a\b\c} =0\ ,\nn\\
&& H_{\a\b \una} = 2i(\C_{\una}Q)_{\a\b}\ \\
&& H_{\a \una \unb} = \left(\C_{\una \unb} Q \chi\right)_{\a} \nn, 
\eea
where
\be
Q=\cases{\C_{11} & (IIA) \cr
              -\s_3& (IIB)}
\ee
and $\chi_{\a}$ is a spinor superfield \cite{nil}
proportional to  the
dilaton superfield of the supergravity background,
ie. $\chi \propto D\phi$.

\subsection{Constant B-field in Flat Spacetime}

We are interested in type II backgrounds with flat
10 dimensional Minkowski spacetime and a constant $B$-field.
In this case, the dilaton is constant and $\chi=0$.
With no distinction between curved and tangent indices,
the torsion constraints \eq{T} are satisfied by the 
flat space pull-back,
\bea\label{E}
&& E_i{}^\umu = \del_i X^\umu - i \thb \C^\umu \del_i \th,
\nn\\
&& E_i{}^\a = \del_i \th^\a .
\eea
Equally, the 3-form constraints \eq{H} are solved by the 
super 2-form $B_{AB}$ with the non-zero components
\bea \label{B}
 && B_{\umu\unu}= {\rm constant}\,,\nn\\
 && B_{\umu \a} = i (\C_\umu Q \th)_\a   
 +i B_{\umu\unu}(\C^\unu \th )_\a\,,\\
 && B_{\a \b} =  - ( \C_\umu \th )_{(\a} (\C^\umu Q \th )_{\b )}
 - B_{\umu\unu} (\C^\umu \th)_\a  (\C^\unu \th)_\b\, .\nn
\eea

Substituting the expressions \eq{E} and \eq{B}  
into the action \eq{action}, we find 
\bea \nn
I = -\frac{1}{2\pi\a'}\int d^2 \xi & \{\sqrt{-g} g^{ij} 
\Pi_i{}^\umu \Pi_j{}^\unu \eta_{\umu\unu} 
+2 i \e^{ij} \del_i X^\umu 
( \thb^1 \C_\umu \del_j \th^1 -\thb^2 \C_\umu \del_j \th^2) \\
&- 2 \e^{ij} (\thb^1 \C^\umu\del_i\th^1)(\thb^2 \C_\umu \del_j \th^2)
+ \e^{ij} \del_i X^\umu \del_j X^\unu B_{\umu\unu}\}, 
\label{flat}
\eea
where we have introduced the more common notation
$(E_i{}^\umu \ \rightarrow )\
\Pi_i{}^\umu = \del_i X^\umu -i\thb^A\C\th^A$ 
with $\th^A$ ($A=1,2$) the two Majorana-Weyl spinors of type II.
We have also used in the above that $\C_{11} = -1$ 
for $\th^1$ (+1 for $\th^2)$ in IIA and
$\s_3 = 1 $ for $\th^1$ (-1 for $\th^2)$ in IIB.
Since \eq{E} and \eq{B} satisfy the constraints 
\eq{T} and \eq{H}, the action \eq{flat} 
is guaranteed to be $\k$-symmetric. 

Thus, we see that for a constant 
$B$-field, the only effect on the action is the addition of the 
last term in \eq{flat}, the usual bosonic coupling to the target 
space. A coupling to the fermionic spacetime
variables $\theta^A$ would arise only if $B$ were not closed.
The couplings to boundary fields can be discussed
using the D-brane vertex operators.

\subsection{D-Brane Boundary Conditions}

We split the 10d indices $\umu =0,1, \cdots, 9$, 
into $\mu =0,1,\cdots, p$ for the directions 
longitudinal to the $Dp$-brane world-volume
\footnote{For convenience in the light-cone gauge,
in section 3 we will take the $9$-direction to be always 
longitudinal to the D-brane.}
and $\mu' =p+1, \cdots,9$ for its transverse directions.
The components $B_{\m\n}$ along the direction of the brane 
are non-zero in the chosen gauge $A=0$. 
The remaining components,
$B_{\umu \nu'}$, are set to zero by a gauge transformation.  
Since  $B_{\m\n}$ is constant, the equations of motion
take the same form as in the case with zero $B$-field,
\bea
& \Pi _i \cdot \Pi _j - {1\over 2} g_{ij} g^{kl}
\Pi _{k } \cdot \Pi _{l } = 0\ ,\label{eom1} \nn\\
& \C \cdot \Pi _{i} P^{i }_{- j } g^{jk }\del_{k} \th^1 = 0 , 
\label{eom2}\nn\\
& \C \cdot \Pi _{i} P^{i }_{+ j } g^{jk }\del_{k} \th^2 = 0 ,
\label{eom3} \label{coveom}\\
& \del_{i}\{\sqrt{-g} g^{ij}\del_{j } X^{\mu }
- 2i P^{\ i } _{-\ j} g^{jk} \bar{\theta }^{1}
\C ^{\mu }\del_{k}\theta ^{1}
- 2i P^{\ i } _{+ \ j} g^{jk} \bar{\theta }^{2}
\C ^{\mu }\del_{k}\theta ^{2} \} =0 , \label{eom4} \nn
\eea
where $P_\pm{}^i_j = \frac{1}{2} 
(\d^i_j \pm \frac{\e^{ik} g_{kj}}{\sqrt{-g}} )$.

For the boundary terms, however, the $B$ field matters.
In the world-sheet conformal gauge, with coordinates
$\xi^i =\{\tau,\sigma\}$, 
the variation of the action \eq{flat} 
requires the following boundary contributions to be zero: 
\bea\label{BC}
&\d X_\umu \left(\Pi_\s{}^\umu -i\thb^1\C_\umu\del_\t\th^1
+i\thb^2\C_\umu\del_\t\th^2 \right) 
+ \d X_\mu \del_\t X^\nu B_\nu{}^\mu 
+i(\thb^A\C_\umu\d\th^A) \Pi_\s{}^\umu \nn \\
& + i\del_\t X^\umu(\thb^1\C_\umu\d\th^1 -\thb^2\C_\umu\d \th^2)
+ ( \thb^1 \C^\umu \d \th^1 \thb^2 \C_\umu \del_\t \th^2 
   -\thb^2 \C^\umu \d \th^2 \thb^1 \C_\umu \del_\t \th^1)|_{\s=0,\pi}  
=0.
\eea
The $\d X$ terms and the $\d \th$ terms must vanish independently. 
Since $\d X^\mu =\d\t\del_\t X^\mu\not=0$, the
vanishing of all the $\d X_\umu$ terms requires the following boundary
conditions ($\s=0,\pi$):
\bea 
\del_\t X^{\mu'}&=& 0, \ \quad \mu'=p+1,...,9 \,,\label{dbc1} \\
\del_\s X^\mu + \del_\tau X^\nu B_\nu{}^\mu &=&0,
\label{nbc1}\ \quad \mu=0,...,p \,,
\\
\thb^1\C^\mu (\del_\s +\del_\t)\th^1 
+\thb^2\C^\mu (\del_\s -\del_\t)\th^2 &=&0 \,.
\label{nbc3}
\eea
We will comment on the last condition later.
Using \eq{dbc1} and \eq{nbc1}, the $\d \th$ terms reduce to
($\s=0,\pi$)
\bea \label{thetaterms}
& i \del_\s X^{\mu'} ( \thb^1 \C_{\mu'} \d \th^1 +  \thb^2 \C_{\mu'} \d
\th^2 ) 
+ i \del_\t X^{\mu} [ (1-B)_{\mu\nu}\thb^2\C^\nu\d\th^2 
- (1+B)_{\mu\nu}\thb^1\C^\nu\d\th^1 ] \nn\\
& + \thb^2 \C_{\umu} \d \th^2  \thb^2 \C^{\umu} \del_\t \th^2 
-\thb^1 \C_{\umu} \d \th^1  \thb^1 \C^{\umu} \del_\t \th^1 . 
\eea
This can be made to vanish if we 
relate the two Majorana-Weyl spinors at the ends of the open 
string, by imposing
\be\label{G}
\th^2 =\C_B \th^1 \quad {\rm at} \ \s=0,\pi ,
\ee
where $\C_B$ is the matrix 
\be \label{aga}
\C_B=
\cases{e^{({1\over2}Y_{\m\n} \C^{\m\n} \C_{11})}
(\C_{11})^{p-2\over 2} \C^{0 \cdots p}& IIA\cr
e^{(-{1\over2} Y_{\m\n} \C^{\m\n} \s_3)}(\s_3)^{p-3\over 2} 
i \s_2 \,\C^{0 \cdots p} \ &IIB} 
\ee
with 
\be
Y=\frac{1}{2}\ln \left(\frac{1+B}{1-B}\right). 
\ee

Indeed, let $C$ be the charge conjugation 
matrix $C$ defined by ${\C^\mu}^T = -C \C^\mu C^{-1}$. The conjugate
spinor  $\thb=\th^T C$ then satisfies $\thb^2 = \thb^1 \Ch$ at $\s=0,\pi$,
where $\Ch_B= C^{-1} \C^T_B C$. 
It is easy to verify that 
\be \label{R1}
 \Ch_B \C^{\mu'} = - \C^{\mu'} \C^{-1}_B 
\ee
and
\be \label{R2}
 \Ch_B \C^\mu = 
\cases{ ( e^{2 Y \C_{11} })^\mu{}_\nu \C^\nu \C^{-1}_B, 
\quad\mbox{IIA}, \cr
        ( e^{ 2 Y \s_3  })^\mu{}_\nu \C^\nu \C^{-1}_B, 
\quad\mbox{IIB}.}
\ee
In particular, since 
$\C_{11} = +1$ for $\th^2$ (-1 for $\th^1)$ in IIA and
$-\s_3 = +1 $ for $\th^2$ (-1 for $\th^1)$ in IIB, we have 
\be \label{R3}
 \Ch_B \C^\mu \th^2 = \left(\frac{1+B}{1-B}\right)^\mu{}_\nu 
\C^\nu \th^1, \quad \s =0, \pi,
\ee
in both IIA and IIB. 
The relations \eq{G}, \eq{R1} and \eq{R3}
guarantee that the first line of \eq{thetaterms}
vanishes. The second line also vanishes by using \eq{R3} twice. 

Observe that with the boundary condition \eq{G}, the action \eq{flat}
is now invariant under the spacetime supersymmetry 
transformations,
\bea \label{susytransf}
&&\d X^\umu = 
i \eb^A \C^\umu \th^A,  \nn\\
&& \d \th^A = \e^A
\eea
only when the constant spinors $\e^A$ are related by
\be\label{econd}
\e^2 = \C_B \e^1\,.
\ee
One can easily verify that $\C_B^2=1$, 
and therefore the previous equation 
completely determines one Majorana-Weyl spinor in terms of the other.
It only leaves 16 independent supersymmetry transformations
linearly realized  on the D-brane.
These are exactly the ones found in \cite{BKOP},
where the matrix $\C_B$ appeared in the $\kappa$-symmetry transformations
for the supersymmetric Born-Infeld action of a D$p$-brane,
in a background where the gauge invariant combination 
${\cal F}=dA-B$ was non vanishing
\footnote{With the slight difference that our $B_{\mu\nu}$ 
is their $-i{\cal F}_{\mu\nu}$ due to our 
normalization of the $B$-field.}. 
Here we derived the same condition from the open string. 
We also note that, for the particular case of the D3-brane, 
$\e^2 = -i \; \mbox{sign(Pf}(B)) \e^1$ when $B \rightarrow \infty$. 
The additional factor of $i$ we have, as compared to \cite{SW}, 
is due to the Minkowskian metric.  

One can also check that the boundary conditions 
are compatible with the supersymmetry \eq{susytransf},
\eq{econd}. Supersymmetric variation of \eq{dbc1} gives
\be
\d X^{\mu'} = i \eb^A \C^{\mu'} \th^A =0 
\ee
using \eq{R1};
while that of \eq{nbc1} gives
\be
\d (\del_\s X^\mu + \del_\tau X^\nu B_\nu{}^\mu) = 
-2i \eb^1 (B/(1-B))^\mu{}_\nu \C^\nu (\del_\s + \del_\t) \th^1 , 
\ee
where we have used $\del_\s \th^2 = - \C_B \del_\s \th^1$
which follows from \eq{G} and the equations of motion \eq{coveom}.
Hence, the boundary conditions are supersymmetric if we also impose
 \be \label{lceq1}
( \del_\s  + \del_\t) \th^1 =0, \quad \s =0, \pi .
\ee 
This, together with \eq{nbc3}, gives 
\be \label{lceq2} 
(\del_\s  - \del_\t) \th^2 =0, \quad \s =0, \pi .
\ee 
Equations \eq{lceq1},\eq{lceq2} are
nothing but part of the equations of motion in light-cone gauge
and thus imposing these boundary conditions is consistent with  
light-cone gauge fixing.

\bigskip

\section{Light-Cone Gauge}

The covariant quantization of the Green-Schwarz superstring
encounters serious obstacles.
On the other hand,
the quantization in light-cone gauge is as straightforward
as in the RNS formalism, with the advantage 
that spacetime supersymmetry is manifest.

Taking the directions 0 and 9 to be along the D-brane
\footnote{
From now on, our discussion will only apply for  $p>0$.},
the light-cone gauge is defined by
\bea
& X^+ (\s,\t) = x^+ +  \a' p^+ \t, \label{lc1}\\
& \C^+ \th^{A} = 0, \,\quad A=1,2,   \label{lc2} 
\eea
where  $X^\pm \equiv \frac{1}{\sqrt{2}} (X^0 \pm X^9)$ and 
$\C^{\pm} \equiv \frac{1}{\sqrt 2} (\C^0 \pm \C^9)$.
Observe that \eq{lc2} is compatible
with the boundary condition \eq{G},
assuming that $B_{0\mu}=B_{9\mu}=0$.

\subsection{Quantization}

In light-cone gauge, only an $SO(8)$ symmetry is apparent.
The surviving components of $\th$ 
\be
S^A= \sqrt{p^+}\C^-\th^A
\ee
are in the two inequivalent (same) 
eight-dimensional spinorial representations
of spin$(8)$ for type IIA (IIB), respectively.
Our ten-dimensional  $\C$-matrices 
satisfy $\{\C^\mu,\C^\nu\} = 2\eta^{\mu\nu}$,
with $\eta^{\mu\nu}={\rm diag}(-1,1,...,1)$. We consider the 
particular realization ($I=1,...,8$)
\bea
&&\C^0 = \left(\begin{array}{cc}
0 & -1_{16} \\ 1_{16} & 0 \\
\end{array} \right), \quad 
\C^I =\left(\begin{array}{cc}
0 & \hat\c^I \\ \hat\c^I & 0 \\ 
\end{array} \right), \quad
\C^9 =\left(\begin{array}{cc}
0 & J \\ J & 0 \\ 
\end{array} \right), \quad \nn\\
&& \C^{11} =\left(\begin{array}{cc}
1_{16} & 0 \\ 0 &  -1_{16} \\ 
\end{array} \right), \quad 
\hat\c^I = \left(\begin{array}{cc}
0 & \c^I_{a\dot a} \\ \c^I_{\dot a a} & 0 \\
\end{array} \right), \quad\quad
J=\left(\begin{array}{cc}
1_8 & 0 \\ 0 & -1_8 \\
\end{array} \right),
\eea
where $\c^I_{a\dot a}$ are the Clebsch-Gordan coefficients
for the coupling of the three inequivalent eight-dimensional 
representations of spin$(8)$ \cite{GS}. 
In this basis, the Majorana-Weyl spinor $\theta^1$ is given by
\footnote{
To avoid possible confusion, we will adopt the notation of putting a
parenthesis around the $SO(2)$ indices $A$, when we 
want to refer to the eight-dimensional real spinors, $S^{(A)}$, 
of spin(8).}
\be
\th^{1} =- \frac{1}{\sqrt{2 p^+}}\left(
\begin{array}{c}
0 \\ 0 \\ S^{(1) a} \\ 0 \\
\end{array}\right).
\ee
Then, the fermion boundary condition (\ref{G}) becomes
\be\label{sbc}
S^{(2)} = N \cdot S^{(1)} \ \quad {\rm at} \ \ \s=0,\pi \,,
\ee
where 
\be
N_{\dot a b} = 
(e^{\frac{1}{2}Y_{ij}\gamma^{ij}} \gamma^{p\cdots 8})_{\dot a b}\,,
\quad\mbox{for IIA}
\ee
and
\be \label{N}
N_{a b} =  
(e^{\frac{1}{2}Y_{ij}\gamma^{ij}} \gamma^{p\cdots 8})_{a b}\,,
\quad\mbox{for IIB} \,.
\ee
In deriving these relations, we have taken
advantage of $\Gamma_{11} \theta^1 = -\theta^1$,
in both type IIA and IIB, to insert $\Gamma_{11}$ on the
right hand side of $\Gamma_B$ in (\ref{G}) in order to convert 
$\Gamma^{01 \cdots (p-1) 9}$ to $\Gamma^{p \cdots 8}$.

The $SO(8)$ triality allows us to clarify the meaning of $N$.
The boundary conditions for the bosonic fields
can be written as usual,
\be\label{xbc}
\del X^I =M^{IJ} \delb X^J \quad {\rm at}\ \s=0,\pi\,,
\ee
with $M^{IJ} = (M^{ij},M^{i'j'} )$ given by 
\be
M^{ij} = \left(\frac{1-B}{1+B}\right)^{ij}, \quad  i=1,\cdots, p-1
\quad\quad \mbox{and} \quad
M^{i'j'} = -\delta^{i'j'}, \quad i' = p, \cdots ,8.
\ee
$M^{IJ}$ is an $O(8)$ orthogonal
matrix acting on the ${\bf 8_v}$ representation.
The eight-dimensional orthogonal matrix $N_{ab}$
is the corresponding matrix acting on the
${\bf 8_s}$ representation of $S^{(1)}$. 
Let $ N_{\dot{a}\dot{b}}=  (e^{\frac{1}{2}Y_{ij}\gamma^{ij}} 
\gamma^{p\cdots 8})_{\dot{a} \dot{b}},$ be the one
for the ${\bf 8_c}$ representation. 
They follow
\be
 \c^I_{a\dot{a}}M^{IJ} = N_{ab} N_{\dot{a}\dot{b}} 
\c^J_ {b\dot{b}}\,.
\ee
A similar relation can be written for IIA. 
Also, it is easy to verify that $N$ is orthogonal, 
\be
N_{ab}N_{ac} = \d_{bc}\,, \quad 
N_{\dot{a}b} N_{\dot{a}c} = \d_{bc}\,.
\ee

In light-cone gauge, the superstring action \eq{flat}
takes the very simple form
\bea \label{lcaction}
I_{l.c.} = \frac{1}{2 \pi\a'}
\int d\s d\t \, \{ -(\del_\t X)^2 +(\del_\s X)^2
+ 2\del_\t X^i \del_\s X^j B_{ij} \nn\\
+ i \a' S^{(1)}\delb S^{(1)} + i \a' S^{(2)}\del S^{(2)}\} \,, 
\eea
where we have introduced the left-moving and right-moving
derivatives $\del = \frac{1}{2}(\del_\t -\del_\s)$ and
$\delb = \frac{1}{2}(\del_\t + \del_\s)$.
The equations of motion \eq{coveom} become the free equations  
\be
\del\delb X^I =0, \quad
\delb S^{(1)} =0, \quad  \del S^{(2)} =0.
\label{lceom}
\ee

With the boundary conditions \eq{sbc} and \eq{xbc}, 
the mode expansions of the fields are
\footnote{The open string ends on the same D-brane.}
\bea
X^i(\t,\s) &=& x_0^i +\a' (p^i_0 \t -p^j_0B_j{}^i\s)\nn\\
&&+\sqrt{\a'} \sum_{n\not=0}\frac{e^{-in\t}}{n}
\left(i a^i_n {\rm cos}(n\s) -a^j_n B_j{}^i\,{\rm sin}(n\s) \right)
\,,\quad i=1,\cdots,p-1,\\
X^{i'}(\t,\s) &=& x_0^{i'}+\sqrt{\a'} \sum_{n\not=0}\frac{e^{-in\t}}{n}
a^{i'}_n  {\rm sin}(n\s) \,,\quad i'=p,\cdots,8, \\
S^{(1)}(\t-\s) &=& \sum_n S_n e^{-i n (\t-\s)},\nn\\
S^{(2)}(\t+\s) &=&  \sum_n N S_n e^{-i n (\t+\s)}.
\eea
The Hamiltonian of the string in light-cone gauge is
\bea \label{lcH}
H &=&\frac{1}{2\pi p^+\a^{'2}}\int_0^\pi d\s\,\{(\del_\t X)^2 
+(\del_\s X)^2 - i \a' S^{(1)}\del_\s S^{(1)} 
+ i \a' S^{(2)}\del_\s S^{(2)} \} \nn \\
&=&\frac{1}{2p^+} \left( p^i p^i 
+\frac{1}{\a'}\sum_{n\not=0}( \a_n^i \a_{-n}^j +\a_n^{i'}\a_{-n}^{i'}
+ 2 n S_n^a S_{-n}^a)\right), 
\eea
where $B^2_{ij} =B_i{}^k B_{kj}$, and we have introduced
$p^i = p_0^j (1-B)_j{}^i$, $\a_n^i = a_n^j (1-B)_j{}^i$, 
$\a_n^{i'} = a_n^{i'}$. The transverse momentum is zero and does
not appear in the Hamiltonian. 

Following the methods of \cite{CH1}, for those modes living on 
the world-volume of the D-brane,
one obtains the (anti) commutation relations 
\bea
&&[x_0^i ,p^j ] =  2 i \left(\frac{1}{1+ B}\right)^{ij}, 
\quad [p^i,p^j]=0, \label{cr1}\\
&&[x_0^i,x_0^j] = 2\pi i \a'\left(\frac{B}{1-B^2}\right)^{ij}, 
\label{x0cr}\\
&&[\a_n^i,\a_m^j] = 2  n \d^{ij} \d_{n+m}, \\
&&\{S^a_n, S^b_m\} = \delta^{ab}\delta_{n+m}, \label{Scr}\\
&&[x_0^i,\a_n^j] = [x_0^i, S_n^a] =[\a_n^i,S_m^a]=0. \label{cr5} 
\eea
The modification in the bosonic modes, $x_0^i$ and $\a_n^i$, 
introduces the notion of non-commutative space for the world-volume
of the D-brane, since one can verify that 
\footnote{The same conclusion is reached if one follows 
the Dirac constrained quantization \cite{CH2}.}
\be\label{commxx}
[X^i(\t,\s),X^j(\t,\s')] = \left\{
\begin{array}{lll}
2\pi i \a'\left(\frac{B}{1-B^2}\right)^{ij}, \quad \s=\s' =0 \cr 
-2\pi i \a'\left(\frac{B}{1-B^2}\right)^{ij},\quad \s=\s' =\pi \cr 
0, \quad  \mbox{otherwise} \,.
\end{array}
\right. 
\ee
But the remaining commutation relations, involving the  
fermion operators $S_n$, remain \emph{unmodified}. 
We will discuss the consequences of this 
for the construction of superspace in non-commutative geometry
in section 4.

\subsection{Unbroken Supersymmetry}

In this subsection,
we will give a more explicit  analysis of the
unbroken supersymmetry living on the brane when $B\not=0$.
We will first identify the unbroken supercharges and
then derive their supersymmetry algebra from the commutation 
relations \eq{cr1}-\eq{cr5}. 

To preserve light-cone gauge conditions \eq{lc1} and \eq{lc2}
under the supersymmetry transformations \eq{susytransf},
we have to perform a compensating $\kappa$-symmetry transformation. 
In order to present concrete formulae, 
we consider the case of type IIB and
denote the supersymmetry parameters
(with the convention that both have negative chirality, 
$\e^A = -\Gamma^{11} \e^A$) by
\be
\e^{A} = \left(
\begin{array}{c}
0 \\ 0 \\ \eta^{(A)a} \\ {\e}^{(A){\dot a}} \\
\end{array}\right)\,.
\ee
The spacetime supersymmetry transformations \eq{susytransf} 
in light-cone gauge read
\bea
&&\d S^{(1)a} = \sqrt{2p^+}\eta^{(1)a} +
\frac{1}{\a'\sqrt{p^+}}\del X^I \c^I_{a\dot{a}}{\e}^{(1)\dot{a}}, \\
&&\d S^{(2)a} = \sqrt{2p^+}\eta^{(2)a} +
\frac{1}{\a'\sqrt{p^+}}\delb X^I \c^I_{a\dot{a}}{\e}^{(2)\dot{a}}, \\
&&\d X^I = \frac{i}{\sqrt{p^+}}S^{(A)a}\c^I_{a\dot{a}}{\e}^{(A)\dot{a}}.
\eea
The Noether supercharges for these symmetries are
\bea
&&Q^{(A)a} = \sqrt{2 p^+} \int_0^\pi \frac{d\s}{\pi} S^{(A)a}, \\
&& Q^{(1)\dot{a}} = \frac{1}{\a'\sqrt{p^+}} 
\int_0^\pi \frac{d\s}{\pi} \del X^I S^{(1)a}\c^I_{a\dot{a}}, \\
&& Q^{(2)\dot{a}} = \frac{1}{\a'\sqrt{p^+}} 
\int_0^\pi \frac{d\s}{\pi} \delb X^I S^{(2)a}\c^I_{a\dot{a}} 
\eea
and they satisfy $Q^{(2)}=N Q^{(1)}$.
The sixteen unbroken 
supercharges are generated by
the linear combinations
\be \label{linear}
Q^a := \frac{1}{2} (Q^{(1)a} + N^{ba} Q^{(2)b}), \quad
Q^{\dot a} := \frac{1}{2} (Q^{(1)\dot{a}} + 
N^{\dot{b}\dot{a}} Q^{(2)\dot{b}}).
\ee
Explicitly,
\bea\nn
Q^a &=& Q^{(1)a} =\sqrt{2p^+} \,S^a_0. \\ \label{Qsusy}
Q^{\dot a} &=& Q^{(1)\dot a}= \frac{1}{\sqrt{p^+ \a'}}  
  \sum_{n\neq 0} \{\c^i_{a\dot a}
({S^a_{-n} \a^i_n}  + \sqrt{\a'} S^a_0 p^i ) 
+  \c^{i'}_{a\dot a}  S^a_{-n} \a^{i'}_n   \}. 
\eea
Now, we can use the commutation relations \eq{cr1}-\eq{cr5}
to compute the algebra of the unbroken supercharges.
It takes the standard form
\bea\label{susyalg}
\{Q^a,Q^b\} &=& 2p^+ \d^{ab}, \\
\{Q^a,Q^{\dot a}\} &=& \sqrt{2} \c^i_{a\dot a} p^i, \\
\{Q^{\dot a},Q^{\dot b}\} &=& 2 H \d^{\dot{a}\dot{b}} ,
\eea
with $H$ being light-cone gauge Hamiltonian \eq{lcH}.
This shows that while the linear combination \eq{linear} 
of unbroken supersymmetry
depends on the $B$-field, the form of the supersymmetry algebra 
is not deformed by the $B$-field. We will also discuss the 
consequences of this in section 4.

But before doing that, 
let us comment on how supersymmetry is realized  in 
the RNS formalism. 
The quantization of the RNS fermions 
$\psi^i = (\psi^i_+,\psi^i_-)$ 
was  carried out in \cite{Schom,CH2} and the following
result was obtained \cite{CH2}:
\bea 
&& \{ \psi_+^i(\s),\psi_+^j(\s') \} = \pi  \eta^{ij}\d (\s,\s'), 
\quad \forall \s, \s' \label{DBpsi1} \\ 
&& \{ \psi_-^i(\s),\psi_-^j(\s') \} = \pi  \eta^{ij}\d (\s,\s'), 
\label{DBpsi2} \\
&& \{  \psi_+^i(\s),\psi_-^j(\s') \} = 
\pi ( (1 + B)(1-B)^{-1} )^{ij}\d (\s,\s'), \quad \s,
\s'\;  \mbox{on the boundary}\,, \label{DBpsi3}
\eea
with the boundary condition
\be
\psi_+^j (1 + B)_j{}^i =  \psi_-^j (1 - B)_j{}^i , \quad
i,j =0,1,\cdots, p. \label{fBC1}.
\ee
Note in particular that the $++$ and $--$ commutation relations are
not modified. 
The boundary condition relates the 
left and right-moving modes for each NS and R sector,
\be
\tilde{d^i_n} = d^k_n  \big( \frac{1 - B}{1+B} \big)_k{}^i,
\quad
\tilde{b^i_r} = b^k_r  \big( \frac{1 - B}{1+B} \big)_k{}^i
\ee
and shows that \eq{DBpsi1} implies \eq{DBpsi2} and 
\eq{DBpsi3}.
The zero modes of \eq{DBpsi1} in the Ramond sector 
form the \emph{usual} 
Clifford algebra acting on the ground state and represent the 
unbroken supersymmetry on the D-brane.

\subsection{Moyal Product}

In this paper we use the oscillator quantization method 
to obtain our results. But the same conclusions are obtained
if one uses conformal field theory methods.
As was derived in \cite{Callan}, the world-sheet propagator 
for the bosonic fields $X$ acquires a crucial new logarithmic 
term in the presence of a constant $B$-field. 
In the Sen-Seiberg limit, it is 
\be
\langle x^i(\t) x^j(0)\rangle = i\frac{\Theta^{ij}}{2} {\rm sign}(\t),
\ee
which by the radial ordering prescription, gives
\be
[x^i(\t),x^j(\t)] = T\left(x^i(\t^+)x^j(\t) - x^i(\t)x^j(\t^+)\right)
= i\Theta^{ij} .
\ee
The leading OPE for normal ordered (i.e. finite) bosonic open 
string vertex operators 
was shown \cite{Schom,SW} to become the Moyal product.

On the other hand, since the fermion propagators satisfy 
first order differential equations, they do not carry any logarithmic
branch cut. It is true that
their boundary conditions modify their OPE;
for instance, in type IIB:
\bea
&& \hat\th^{(A)a}(\t) \hat\th^{(A)b}(\t') \sim \frac{\d^{ab}}{\t-\t'} ,
\quad A=1,2 \ ({\rm not \ summed})\,,\\
&& \hat\th^{(1)a}(\t) \hat\th^{(2)b}(\t') \sim \frac{N^{ab}}{\t-\t'}\,.
\eea
However, there is no modification of their anti-commutators,
\be
\{\hat\th^{(A)a},\hat\th^{(B)b}\} = 
T\left(\hat\th^{(A)a}(\t^+) \hat\th^{(B)b}(\t) 
+ \hat\th^{(A)a}(\t) \hat\th^{(B)b}(\t^+) \right) = 0
\ee
which agrees with \eq{Scr}.

Let us remark on how the Moyal phases come out 
in the oscillators quantization method.
Consider the zero momentum Fock-space ground state of the 
open bosonic string, $|0,0\rangle$. The tachyon state 
$|0,k\rangle$ is obtained by the insertion in the far past of the
tachyon vertex operator, located at the world-sheet boundary 
(take for instance $\s=0$) 
\be
V_0(k;\t) = : e^{ik\cdot X(0,\tau)}: .
\ee
Explicitly, the state is 
\be\label{0,k}
|0,k\rangle = \lim_{\t\to +i\infty} e^{-i\t} V_0(k;\t) |0,0\rangle . 
\ee
Since $V_0(k,\t)$ is normal ordered, (\ref{0,k}) is just
\be\label{def}
|0,k\rangle = e^{ik\cdot x_0} |0,0\rangle .
\ee
Observe that (\ref{def}) satisfies the intuitive property
that in the absence of a $B$ field, 
\be\label{consis}
|0,k_1+k_2\rangle = e^{i(k_1+k_2)\cdot x_0}|0,0\rangle
=e^{ik_1\cdot x_0}|0,k_2\rangle = 
e^{ik_2\cdot x_0}|0,k_1\rangle .
\ee
But when there is a constant background $B\not=0$, the state \eq{def} 
with momentum $k$ will no longer satisfy 
the last three equalities in \eq{consis}.  
Instead, the Moyal phase appears due to \eq{x0cr}
\be
\lim_{\t\to +i\infty} e^{-i\t} V_0(k_2,\t) |0,k_1\rangle = 
e^{i k_2\cdot x_0} e^{i k_1 \cdot x_0} |0,0\rangle
= e^{i k_1\wedge k_2} |0,k_1+k_2\rangle,
\ee
where $k_1\wedge k_2 := \frac{1}{2} \Theta^{ij}k_{1,i} k_{2,j}$. 
This explains the appearance of the Moyal phases
in interaction vertices. For example, for the three tachyon vertex
\be
\langle 0,k_1|V_0(k_2,0)|0,k_3\rangle =
\langle 0,k_1| e^{ik_2\wedge k_3} |0,k_2+k_3 \rangle =
e^{ik_2\wedge k_3} \d(-k_1+k_2+k_3) .
\ee

\bigskip

\section{Discussion: Superspace and Superfields}

We have derived that for the non-commutative 
supersymmetric quantum field
theories obtained from string theory, 
the fermionic variables $\theta$ continue to
anti-commute. Motivated by this, we introduce the
non-commutative superfield, $\Phi(x,\th)$, to be a local function 
\be
\Phi(x,\th)=\phi(x) + \th\lambda(x) + \cdots
\ee
Obviously, it
continues to have a finite Taylor expansion in $\theta$. 
The Moyal product for superfields evaluated
on the supermanifold completely ignores the fermionic
coordinates, {\it i.e.},
\bea
\Phi^I(x,\theta)\star \Phi^J(x,\theta) &=& 
e^{\frac{i\Theta^{ij}}{2}\frac{\del}{\del x^i}\frac{\del}{\del y^j}}
\Phi^I(x,\theta)\Phi^J(y,\theta)|_{x=y} \\
&=&\phi^I(x)\star\phi^J(x) + \th \left(\phi^I(x)\star\lambda^J(x)
+\lambda^I(x)\star\phi^J(x)\right) +\cdots
\eea

In order to be concrete, let us focus on the case 
of 4d ${\cal N}=1$ supersymmetry
\footnote{We follow the conventions of \cite{WB}.}.
It has supercharges $\{Q_\a,{\overline Q}_{\dot \a}\}$ and the
supersymmetry algebra can be obtained by a convenient truncation of 
the 10d ${\cal N}=1$ supersymmetry algebra \eq{susyalg}
\footnote{Properly, the algebra \eq{susyalg} 
was derived in light-cone gauge. 
Here we extend its 4d ${\cal N}=1$ subalgebra 
to its covariant off-shell formulation.}
The realization of the 
supercharges as first order differential operators on the superspace 
$\{x,\th,{\bar \th}\}$ is the same as in the ordinary spacetime case.
The same is true for the supersymmetric covariant derivatives,
$\{D_\a,{\overline D}_{\dot \a}\}$.
Supersymmetry transformations are 
generated by the usual shifts of the superspace variables:
\bea
&&\th \to \th + \xi
\\
&&x^\mu \to x^\mu + i\th\s^\mu\bar{\xi} -i\xi\s^\mu\bar{\th} \,,
\eea
and the same sort of supersymmetric field transformations 
are obtained. The only modification comes from
replacing the ordinary commutative product by 
the Moyal product for any product 
of $x$-dependent quantities.

Superfields can be introduced as field representations of the
supersymmetry algebra. For example,  chiral superfields satisfy
${\overline D_{\dot \a}}\Phi =0$. 
Supersymmetric invariant actions are constructed in the same way, as
integrals in superspace of some  $x$-dependent superfield expression.
The most general Lagrangian that can be built from scalar 
chiral superfields takes the form
\be\label{WZ}
{\cal L} = \int d\th^2 d\thb^2 \,{\cal K}({\overline \Phi},\Phi) +
\int d\th^2\, {\cal W}(\Phi) + {\rm h.c.}
\ee
We will still refer to ${\cal K}({\overline \Phi},\Phi)$
as the K$\ddot{\rm a}$hler potential, 
simply because we still have the symmetry
\be
{\cal K}({\overline \Phi},\Phi) \to {\cal K}({\overline \Phi},\Phi)
+ F(\Phi) + {\overline F}({\overline \Phi}). 
\ee
It is easy to show that the auxillary fields are nonpropagating and
can be eliminated to give a supersymmetric action in terms of on-shell
states only. This apparently differs from the situation in 
some higher derviative actions, e.g. in $N=2$ SYM with nonholomorphic
corrections, where the auxillary fields can become propagating. 
\footnote{We thank J.P. Derendinger 
for useful discussions and comments on this.} 

Observe that, for chiral superfields, one still has the 
powerful notion of a holomorphic superpotential ${\cal W}(\Phi)$,
which only requires a chiral $\th$-integral to be included in the 
Lagrangian.  
Also, one can formulate Feynman super-graphs for \eq{WZ}. 
All the quadratic divergences are cancelled out manifestly.
It is easy to show that any perturbative quantum
corrections to the effective action will be an integral over the whole
superspace. Then, modulo the usual subtleties 
arising from massless fields, 
the perturbative non-renormalization of the F-term
still holds for non-commutative supersymmetric field theories.
\footnote{See \cite{mrs,abk} for a recent discussion of IR  
effects in non-commutative field theories.
A $1/p\circ p$ singularity in the two point
function (arising from the 1 loop non-planar diagram)
was found in the  non-commutative theory.
In particular, it was shown that a light mode is generated 
whose mass depends on the coupling of the theory \cite{mrs}.
They also show that  no new pole is  
generated in the supersymmetric case.
This can also 
be verified easily by doing a simple 
Feynman super-graph calculation.}
It has been argued in
\cite{cr} that noncommutative Wess-Zumino model is renormalizable.

Obviously, the most interesting case is 
non-commutative supersymmetric Yang-Mills, where the 
Ward identities of the non-commutative gauge symmetry
and its sensitive embedding in a consistent superstring theory
suggests that it is a well defined quantum theory. 
But we leave this issue for future studies.

In this letter, we have shown that
the manifestly supersymmetric formulation of the
non-commutative supersymmetric field theories,  
obtained from string theory in flat spacetime and 
constant $B$-field backgrounds, 
shares a large number of the features 
in common with the 
ordinary supersymmetric field theories.
The notion and construction of superspace and superfield 
is the same, without additional deformation of the fermionic variables
$\th$. In more general string backgrounds, one may get  a more
non-trivial ``Moyal product'' for the fermionic coordinates
of the relevant superspace. It would  be interesting to know 
which are the admissible ``Moyal products'' for a given supermanifold 
and whether and how they arise in string theory.   

Manifest supersymmetric formulations have proven to 
provide useful and powerful concepts,
such as holomorphy and non-renormalization theorems,
for analyzing different quantum aspects of the 
supersymmetric field theories.
It is quite reasonable that a manifestly supersymmetric 
treatment of non-commutative supersymmetric field theories
will also shed light on the dynamics of these theories.

\bigskip

\section*{Acknowledgments}

We thank Miao Li for
informing us  that some of the issues considered in this paper 
had also been considered by him  and Yong-Shi Wu in a slightly
different approach. 
CSC thanks A. Bilal, J.P. Derendinger, M. Li and R. Russo for
useful discussions and comments. 
FZ thanks J. Distler, M.J. Perry and P. Pouliot 
for helpful discussions and A. Kashani-Poor for a careful 
reading of the manuscript.
We acknowledge the Aspen Center of Physics, where this work 
started, for hospitality.  
The work of CSC is supported by the Swiss National Science
Foundation. The work of FZ is supported by NSF Grant PHY9511632 and
the Robert A.~Welch Foundation.


\end{document}